\pdfoutput=1
\documentclass[twocolumn,showkeys]{aastex631}
\usepackage{epstopdf}
\usepackage{xcolor}
\usepackage{xspace}
\usepackage{graphicx}
\usepackage{amssymb, amsmath}
\usepackage{natbib}
\usepackage{float}
\usepackage{hyperref}
\usepackage{mathptmx}
\usepackage{verbatim}
\usepackage{textcomp}
\usepackage{latexsym}
\usepackage{multirow}
\usepackage{wasysym}
\usepackage{academicons}
\usepackage{gensymb}
\usepackage{longtable}
\usepackage{ifthenx}
\usepackage{comment}
\usepackage{silence}
\shorttitle{Proper Motion of HST-1}
\shortauthors{Thimmappa et al.}

\accepted{2024 April 29}
\revised{2024 April 10} 
\received{2023 October 10} 

\begin{document}

\title{ {\it Chandra} Study of the Proper Motion of HST-1 in the Jet of M87}

\correspondingauthor{R.~Thimmappa}
\email{rameshan.thimmappa@villanova.edu}

\author[0000-0001-5122-8425]{Rameshan Thimmappa}
\affiliation{Villanova University, Department of Physics, Villanova, PA 19085, USA}

\author[0000-0002-8247-786X]{Joey Neilsen}
\affiliation{Villanova University, Department of Physics, Villanova, PA 19085, USA}

\author[0000-0001-6803-2138]{Daryl Haggard} 
\affiliation{McGill University, Montreal, QC, Canada}

\author[0000-0001-6923-1315]{Mike Nowak} 
\affiliation{Washington University in St. Louis, St. Louis, MO 63130, USA}

\author[0000-0001-9564-0876]{Sera Markoff} 
\affiliation{API, University of Amsterdam, Amsterdam, Netherlands}

\begin{abstract}

The radio galaxy M87 is well known for its jet, which features a series of bright knots observable from radio to X-ray wavelengths. We analyze the X-ray image and flux variability of the knot HST-1 in the jet. Our analysis includes all 112 available \textit{Chandra} ACIS-S observations from 2000-2021, with a total exposure time of $\sim$887 ks. We use de-convolved images to study the brightness profile of the X-ray jet and measure the relative separation between the core and HST-1. From 2003-2005 (which coincides with a bright flare from HST-1), we find a correlation between the flux of HST-1 and its offset from the core. In subsequent data, we find a steady increase in this offset, which implies a bulk superluminal motion for HST-1 of 6.6$\pm$0.9 c (2.0$\pm$0.3 pc yr$^{-1}$), in keeping with prior results. We discuss models for the flux-offset correlation that feature either two or four emission regions separated by tens of parsecs. We attribute these results to moving shocks in the jet, that allow us to measure the internal structure of the jet.

\end{abstract}


\section{Introduction} 
\label{sec:intro}
In extragalactic radio sources, relativistic jets travel hundreds of kpc from the central engine \citep{Blandford74, Rees78, Begelman84, Begelman89, Bridle84}. These prominent jets can be seen in high-luminosity objects, such as 3C 273 \citep{Harris87}, PKS 0637 - 752 \citep{Schwartz00}, and Pictor\,A \citep{Wilson01}, as well as nearby objects like Cen\,A \citep{Hardcastle07} and the radio galaxy M87 \citep{Harris03}.

M87 is an FR-I type nearby active galactic nucleus (AGN) located at a distance of $\sim$ 16.7$\pm$0.2 Mpc \citep{Mei07}. It hosts an extragalactic jet \citep[discovered by][]{Curtis18}, which features a series of bright knots visible from radio to X-ray. The X-ray knots correlate with optical knots within the jet \citep{Marshall02}. The cooling time of X-ray emitting electrons (with $\gamma \sim$10$^7$) will be a few years and optical emitting electrons will be a few decades. If the emission is synchrotron, the short cooling timescales in the X-ray require that the X-ray emitting electrons be accelerated locally within the jet \citep{Harris06a}. Thus X-ray emission regions are ideal probes of energetic particle acceleration processes in jets.

\begin{figure*}[ht!]
	\centering 
\includegraphics[width=0.95\textwidth]{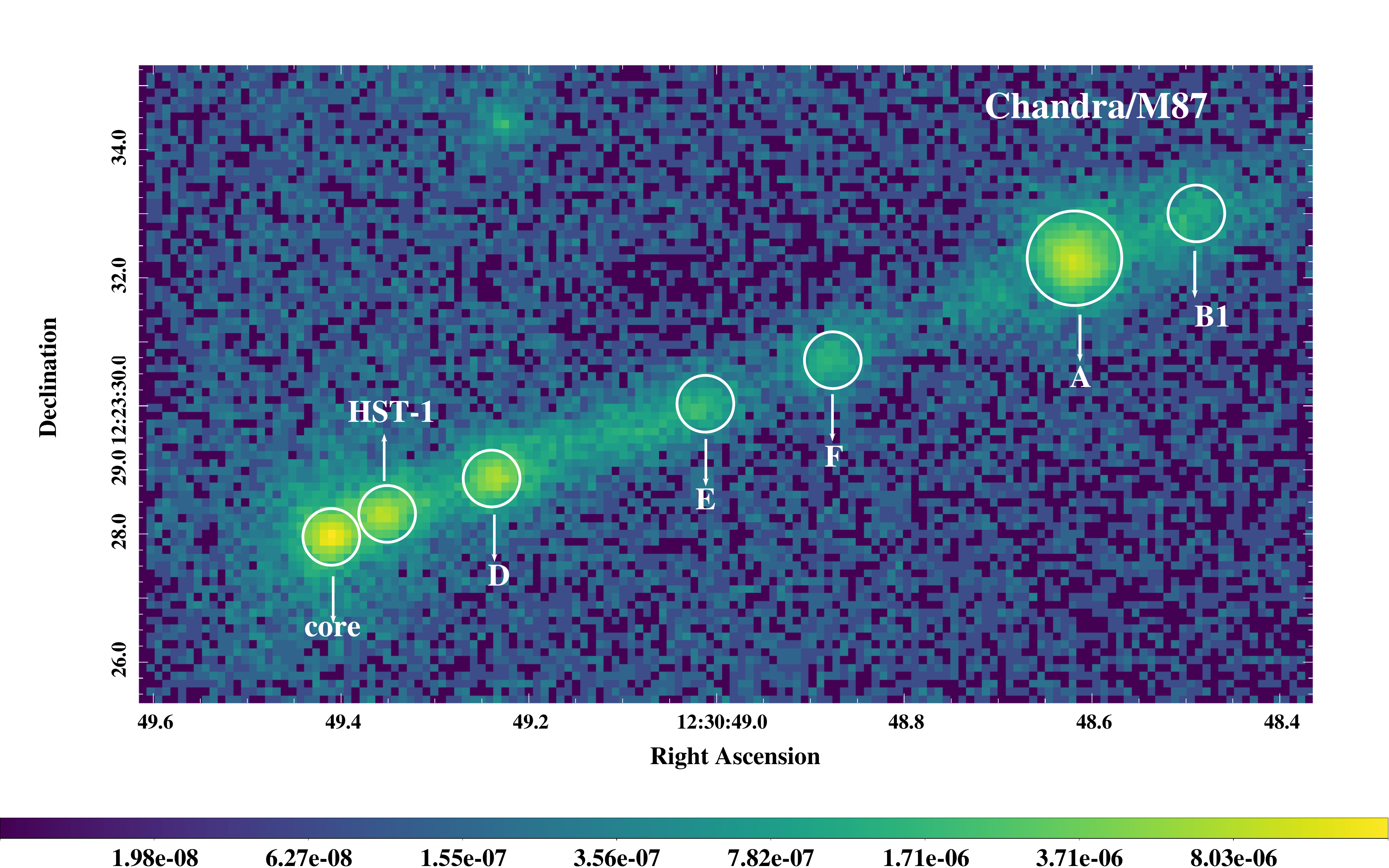}
\caption{The exposure-corrected $0.4-8.0$ keV image of the jet in M87, binned to 0.25 pix resolution for ObsID\,352.} 
\label{fig:M87_jet}
\end{figure*}

One of the most exciting features of the jet in M87 is the inner knot HST-1, observed a projected distance of $\sim$ 70 pc from the active core \citep{Biretta99, Harris03, Harris06, Perlman03}. HST-1 is one of the brightest features in the jet and has been studied in detail using multiple telescopes from radio to optical, near-UV, X-ray, and gamma-ray \citep{Biretta99, Marshall02, Wilson02, Harris09, Madrid09, Chen11, Giroletti12}. In 2003, a bright X-ray flare was detected from HST-1: the source rose in flux until 2005 and then faded until returning to its prior brightness in 2008 \citep{Harris09}. Such flares could provide evidence of localized changes and the existence of shock regions in the jets, which can accelerate particles to high energies \citep{Perlman11, Imazawa21}. Analyzing a large set of \textit{Chandra} observations could therefore trace the physical process and the behavior of the knots in this relativistic jet.   
The proper motion of the jet and its components have also been a major source of interest and have been studied at a range of spatial scales. For example, the MOJAVE (Monitoring Of Jets in Active galactic nuclei with VLBA Experiments) program using the parsec-scale \citep{Homan09, Lister09}, Very Long Baseline Interferometry (VLBI) has shown that the FWHM apparent opening angle is $\sim$6$^{\circ}$.9 out from the radio core \citep{Reid89, Dodson06, Kovalev07, Asada14} on parsec scales, while observations at larger scales have shown that the dynamic structure varies within the flaring X-ray region of the jet \citep{Cheung07, Giroletti12}. Meanwhile, the {\it Hubble} Space Telescope \citep[HST;][] {Biretta99, Madrid09, Meyer13} and the {\it Chandra} X-ray Observatory \citep{Snios19} have measured the proper motion of HST-1 and other knots along the jet over time. These measurements often reveal significant relativistic motion, where jet features move at velocities close to the speed of light. 
 
Here, we investigate the proper motion of HST-1 in the jet of M87 using a large set of \textit{Chandra} X-ray images. We find a complex relationship between the flux of HST-1 and its offset from the core, and we develop a toy model to explain the observed behavior.

This paper is organized as follows. In section 2, we provide the details of the \textit{Chandra} data reduction. In Section 3, we present the result of our imaging study using spectral modeling, point-spread function (PSF)/MARX ray-tracing simulations, image deconvolution, and radial profiles of the jet. In Section 4, we analyze the apparent motion of HST-1 and present our model of the flux-offset correlation. We compare our results to previous measurements of the proper motion of HST-1 and discuss the implications in Section 5.

Throughout this work we define the photon index $\Gamma$ as $F_{\varepsilon} \propto \varepsilon^{-\Gamma}$ for photon flux spectral density $F_{\varepsilon}$ and  photon energy $\varepsilon$; the spectral index is $\alpha = \Gamma - 1$.

\section{Chandra Data} 
\label{sec:data}

HST-1 was observed on-axis with the Advanced CCD Imaging Spectrometer \citep[ACIS-S;][]{Garmire03} onboard the {\it Chandra} X-ray Observatory \citep{Weisskopf00} from 2000 to 2021. An example image is shown in Figure \ref{fig:M87_jet}. For this study\footnote{{\url{https://doi.org/10.25574/cdc.203}}}, we used 112 ACIS-S pointings with a total exposure time of $\sim$ 884 ks. The observations from 2000 and 2018 can be found in \citet{Yang19}, and additional observations are ID 352 (2000), 21457 (2019), 21458 (2019), 23669 (2021), and 23670 (2021). The quality of the data for the focused region varies between the different pointings because the effective PSF is a complicated function of position in the image, the source spectrum, and the exposure time. 

The observational data were reprocessed using the {\ttfamily chandra\_repro} script as per CIAO v4.14 \citep{Fruscione06} analysis threads\footnote{\url{http://cxc.harvard.edu/ciao/threads/}}, using CALDB v4.9.7. Pixel randomization and readout streaks were removed from the data during processing and astrometric correction was performed. Point sources in the vicinity of the jet were detected with the {\ttfamily wavdetect} tool using the minimum PSF method and removed. The {\ttfamily wcs\_match} and {\ttfamily wcs\_update} tools were used to match all the ObsIDs with reference ObsID 352. All the event and aspect solution files were updated subsequently. For our analysis, we selected photons in the range of $0.4-8.0$ keV. Photon counts and spectra were extracted for the source and background regions from individual event files using the {\ttfamily specextract} script. Spectral fitting was done with the {\fontfamily{qcr}\selectfont Sherpa}\footnote{\url{https://cxc.cfa.harvard.edu/sherpa}} package \citep{Freeman01}. The achieved frame time of all reprocessed observations is 0.4\,s \citep{Harris06, Yang19}. The moderate count-rate of $\simeq$ 0.4 counts/pix/s at maximum for HST-1 located in the S3 chip implies a small chance for pile-up in the detector \citep{Davis01}; we included the pile-up model when performing {\fontfamily{qcr}\selectfont MARX} ray-tracing simulations (see section \ref{sec:PSF}).

\section{Data Analysis} 
\label{sec:analysis}

The goal of this work is to use deconvolved images to measure the radial profile of the jet and measure the offset of HST-1 from the core of M87. 
Although some authors have fit models to data directly for other sources, for HST-\,1, we use deconvolved images because it is more straightforward to account for pileup during {\fontfamily{qcr}\selectfont MARX} simulation (Section \ref{sec:PSF}) than with direct fitting. Deconvolved images have been used in imaging analysis to study the structure of astronomical sources, for example, to examine the jet's broadband spectrum and to study the nature of particle acceleration and a correlation between X-ray intensity and optical polarization \citep{Perlman03}, the proper motion of the pulsar PSR J1741-2054 \citep{Auchettl15}, the large-scale jet of 3C 273 \citep{Marchenko17}, and the western hotspot of Pictor\,A \citep{Thimmappa20, Thimmappa22}.  In general, image deconvolution requires a spectrum of the source, modeling of the PSF, and an algorithm to extract the deconvolved images from the data. 
 
\subsection{Spectral Modeling}
\label{sec:spectrum}

\begin{figure}[t!]
\centering 
\includegraphics[width=\columnwidth]{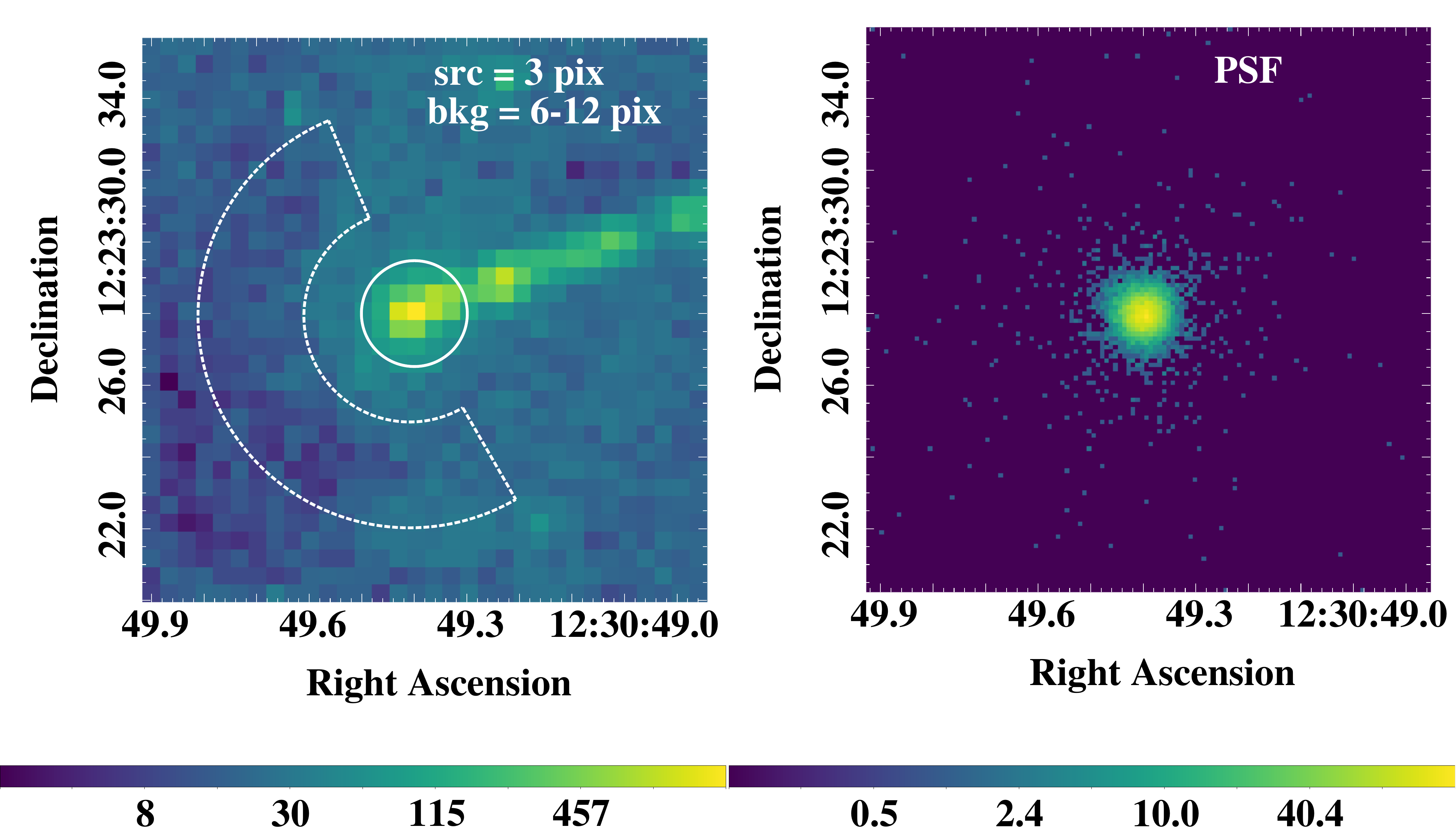}
\caption{\textit{Left panel}: ACIS-S image of the source (core+HST-1) in the jet of the M87 radio galaxy, within the energy range 0.4–8 keV for the ObsID 352 (with 1 px binning). The source extraction region for the spectral modeling is denoted by the solid circle (3 px radius), and the background region by a dashed half annulus (6–12 px). \textit{Right panel}: the PSF simulated at 0.25-pix resolution for ObsID\,352 (as discussed in Section 3.2).} 
\label{fig:src_bkg_PSF}
\end{figure}

For our spectral analysis, we extracted the source spectrum in each ObsID from a circular region (position: RA\,=\,12:30:49.3999, DEC\,=\,+12:23:28.000) with a radius 3 px ($\simeq 1.5^{\prime\prime}$, for the conversion scale $0.492^{\prime\prime}/{\rm px}$), and the background was set as a half-annulus with an inner and outer radius of 6 and 12 px, respectively; we excluded the half of the annulus that is on the same side of the source as the extended X-ray jet, which is shown in Figure \ref{fig:src_bkg_PSF} left panel. The background-subtracted source spectra were fitted assuming a {\fontfamily{qcr}\selectfont power-law + APEC} emission model modified by the Galactic column density $N_{\rm H,\,Gal} = 0.0126 \times 10^{22}$ cm$^{-2}$ \citep{HI4PI2016}. 

A {\fontfamily{qcr}\selectfont Power-law} model with photon index $\Gamma \simeq 2$ and {\fontfamily{qcr}\selectfont vapec} model with a plasma temperature $kT\simeq 0.88$ keV provides a reasonable description of the source spectra. It is sufficient in particular for the purpose of the PSF modeling. We note that analogous fits with the Galactic absorption left free returned similar results, with slightly decreased values of $N_{\rm H,\, Gal}$, $\Gamma$, and $kT$. Finally, including the {\ttfamily jdpileup} model in the fitting procedure does not affect the best-fit values of the model parameters, as the fraction of piled-up events that result in a good grade turns out to be very low.

\subsection{PSF Modeling}
\label{sec:PSF}
To model the {\it Chandra} PSF at the source position (core+HST-1), we used the \textit{Chandra} Ray Tracer ({\fontfamily{qcr}\selectfont ChaRT}) online tool \citep{Carter03}\footnote{\url{http://cxc.harvard.edu/ciao/PSFs/chart2/runchart.html}} and the {\fontfamily{qcr}\selectfont MARX} software \citep{Davis12} \footnote{\url{https://space.mit.edu/cxc/marx}}. For all ObsIDs, the centroid coordinates of the selected source regions were taken as the position. The source spectra used for the {\fontfamily{qcr}\selectfont ChaRT} tool is the {\fontfamily{qcr}\selectfont power-law + vapec} model in the 0.4--8.0 keV band, as described in Section \ref{sec:spectrum}. Since each particular realization of the PSF is different due to random photon fluctuations, in each case a collection of 50 event files was made, with 50 iterations using {\fontfamily{qcr}\selectfont ChaRT} by tracing rays through the {\it Chandra} X-ray optics. The rays were projected onto the detector through {\fontfamily{qcr}\selectfont MARX} simulation, taking into account all the relevant detector effects, including pileup and energy-dependent sub-pixel event repositioning. The PSF images were created with a bin size of 0.25 pixels. An example of the simulated PSF image for ObsID\,352 is presented in Figure \ref{fig:src_bkg_PSF} (right panel).

\begin{figure*}[ht!]
	\centering 
	\includegraphics[width=0.77\textwidth]{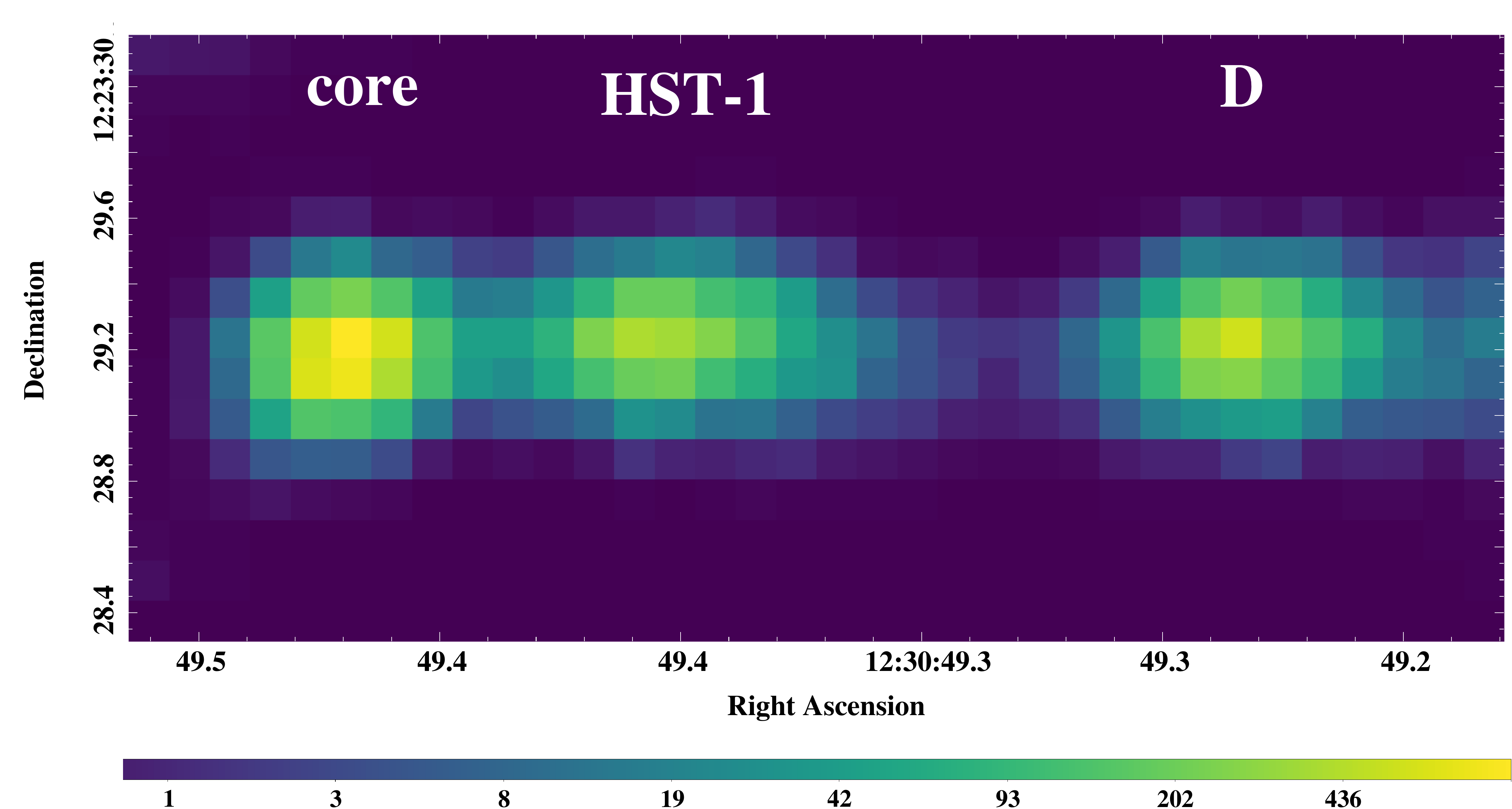}
 	\includegraphics[width=0.77\textwidth]{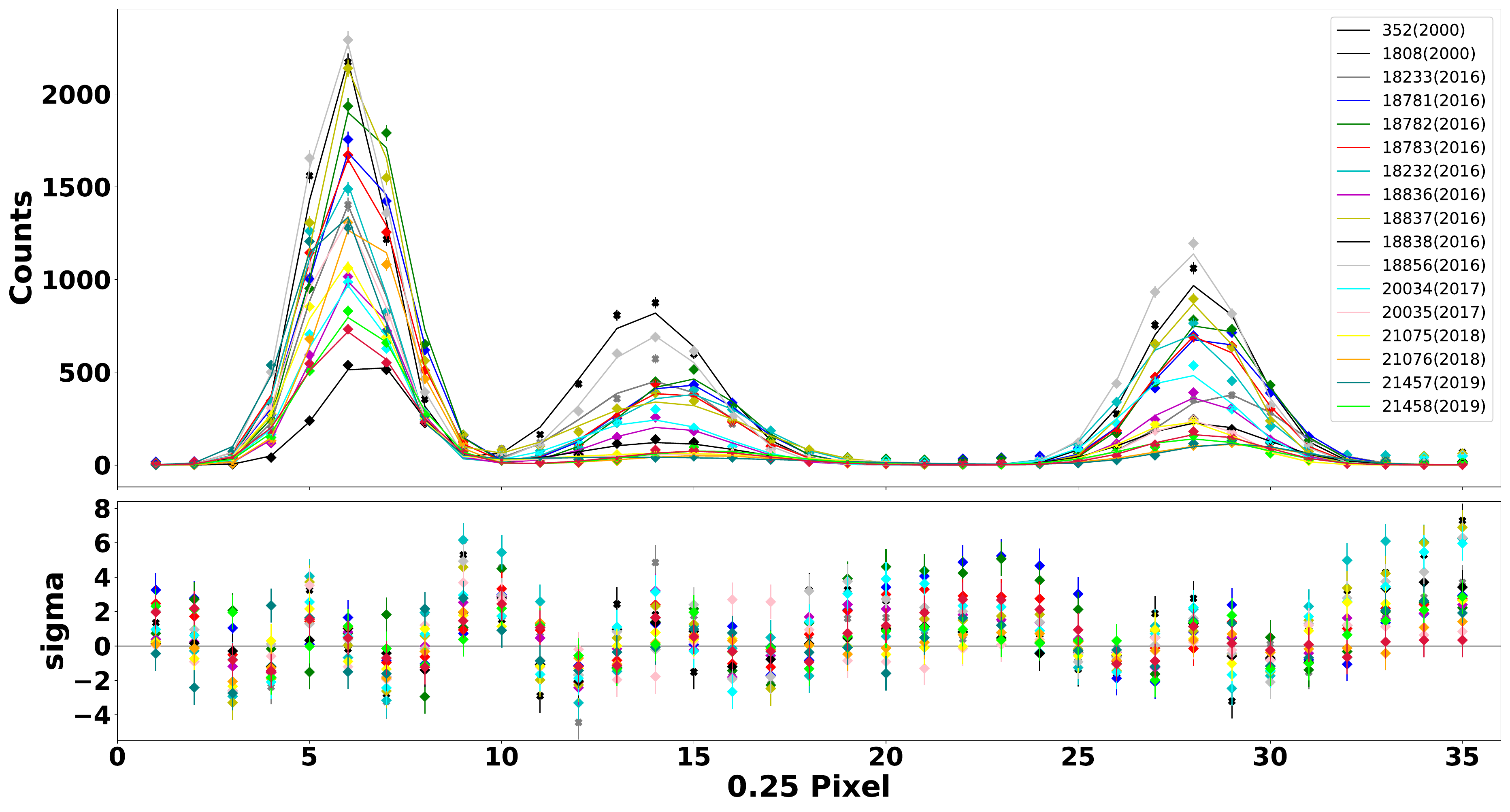}
	\includegraphics[width=0.77\textwidth]{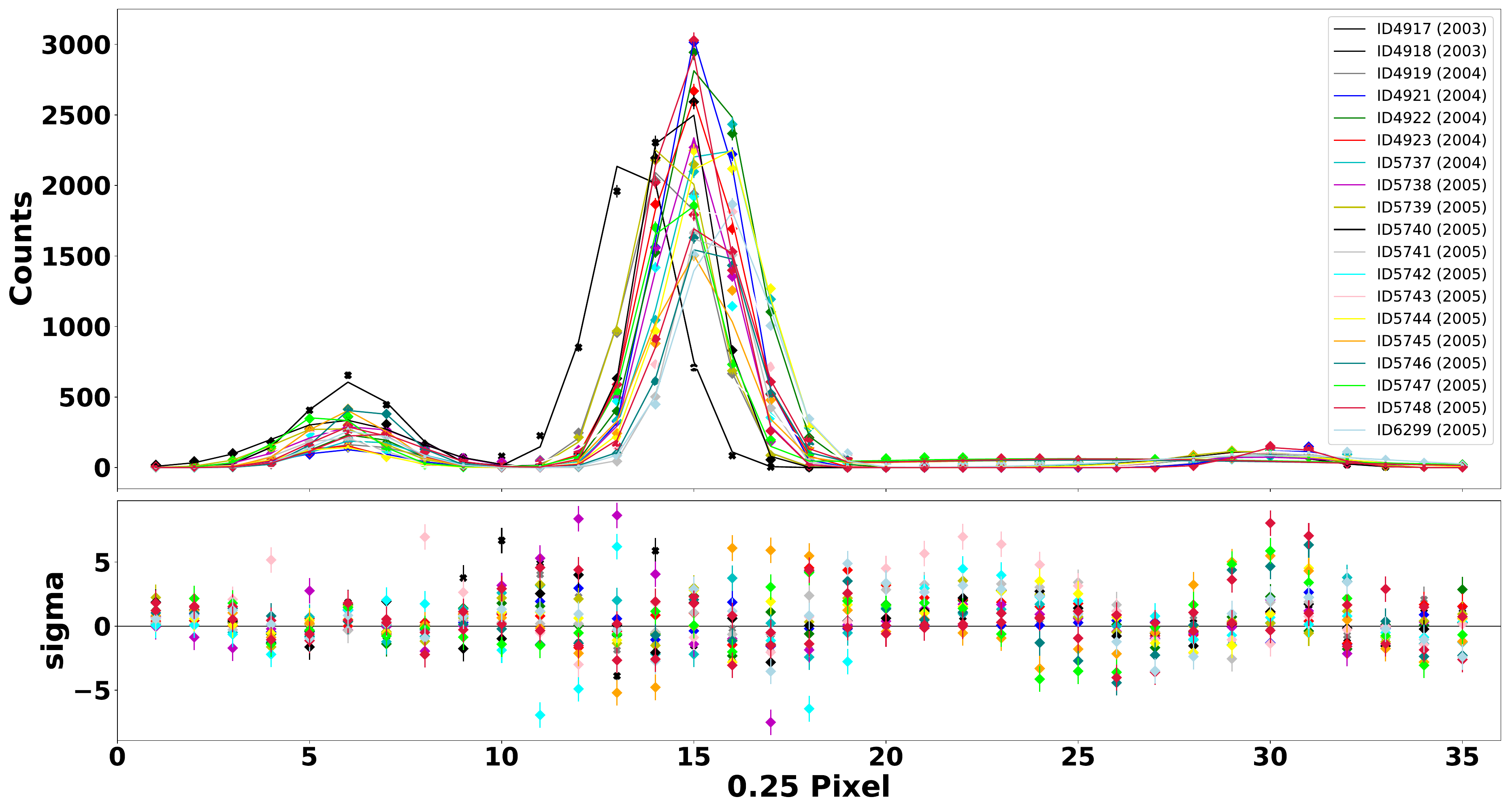}
	\caption{\textit{Top panel}: The $0.4-8.0$ keV energy band exposure-corrected deconvolved image of the jet with 0.25-pix resolution of ObsID\,352. \textit{Middle panel}: The radial profile of long exposure ($>10$ ksec) ObsIDs. \textit{Bottom panel}: The radial profile between 2002-2008 ($<5$ ksec) ObsIDs.} 
\label{fig:radial_profile}
\end{figure*}

\begin{figure*}[ht!]
	\centering 
	\includegraphics[width=0.99\textwidth]{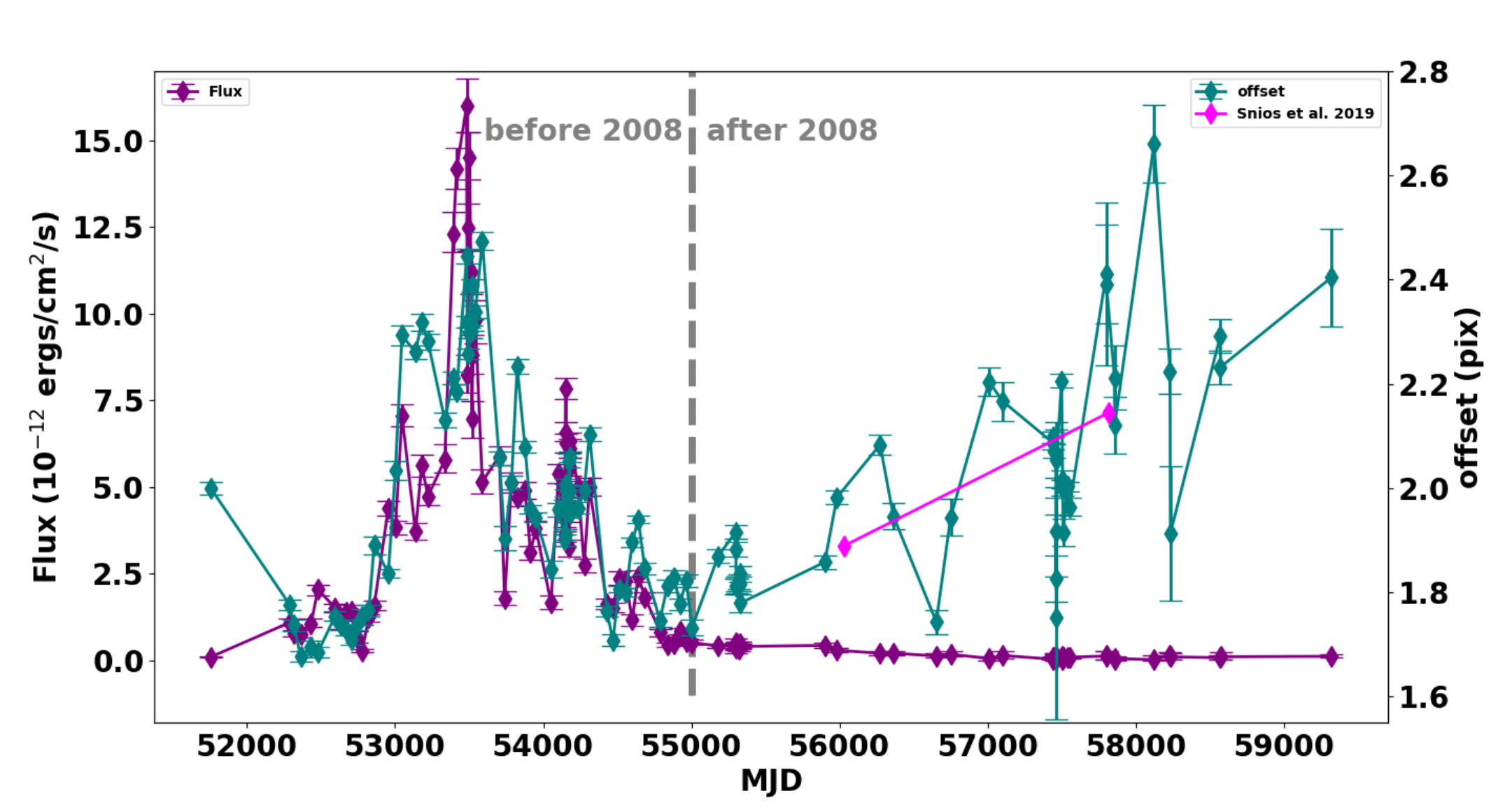}
	\caption{Flux lightcurve and offset of 112 ObsIDs versus time.} 
   \label{fig:HST-1_lightcurve}
\end{figure*}

\subsection{Image Deconvolution}
\label{sec:deconvolve}
We used the Lucy-Richardson Deconvolution Algorithm \citep[LRDA,][]{Lucy74}, which is implemented in the {\fontfamily{qcr}\selectfont CIAO} tool {\ttfamily arestore}, to remove the PSF blurring, and in this way to restore the intrinsic surface brightness distribution of the core and HST-1. This method does not affect the number of counts in the image, but only their distribution. The algorithm requires an image form of the PSF, which is provided by our {\fontfamily{qcr}\selectfont ChaRT} and {\fontfamily{qcr}\selectfont MARX} simulations as described in Section \ref{sec:PSF}, and exposure-corrected maps of the source. We obtained the deconvolved images for each ObsID, with 50 different random realizations of the simulated PSF; those 50 deconvolved images (Figure \ref{fig:radial_profile}, top panel) were then averaged to a single image using {\ttfamily dmimgcalc} tool.

\section{Results}
\subsection{The Radial Profile}
\label{sec:radial_profile}
We use the 0.25-px or quarter-pixel (qp) deconvolved images described in Section \ref{sec:deconvolve} to study the radial profiles of the projected jet. The CIAO {\ttfamily dmregrid2} tool was used to rotate the deconvolved images; We selected a rectangular region $\sim$ 18$\times$4\,pixel$^2$ $\sim$72$\times$16 (0.25)\,pixel$^2$ around the jet,  which includes the core, HST-1 and knot D, as shown in Figure\,\ref{fig:radial_profile}, top panel. The counts were summed for each column of quarter pixels perpendicular to the jet axis to produce the radial profile. We assumed \texttt{chi2gehrels} statistics for the uncertainties. We performed this exercise separately for each ObsID. Based on the counts from deconvolved images we produced the radial profile of the jet, shown in Figure \ref{fig:radial_profile}. The regions of the core, HST-1, and knot D are labeled in the top panel, and the respective radial profiles are shown in the center and bottom panels. The center panel shows the radial profile before and after the flare of HST-1, where observations have longer exposure times ($>$10 ksec). The bottom panel shows the radial profile of HST-1 during the flare between 2003 and 2005, where the HST-1 brightness dominates over the core and exposure times tended to be shorter. 

The position of the core, HST-1, and knot D are measured with one-dimensional Gaussian fitting. The model was fit to each ObsID using {\fontfamily{qcr}\selectfont Sherpa}, with the \texttt{Levenberg-Marquardt} method and \texttt{chi2gehrels} statistic. The uncertainties of the parameters are calculated from the confidence method using the {\ttfamily conf} command in {\fontfamily{qcr}\selectfont Sherpa}, which computes confidence intervals for the given model parameters. \cite{Snios19} noted that it is difficult to achieve the required astrometric accuracy to measure the proper motion of HST-1 with ACIS, so we opted to perform relative astrometry instead, fitting directly for the offset of HST-1 relative to the core.

The flux of HST-1 was measured from 0.4-8 keV using the \textsc{ISIS} software package \citep{Houck02}. For each observation, we model the HST-1 source and background spectra jointly: the background spectrum with a \texttt{vapec} component and the knot spectrum as the sum of a \texttt{pegpwrlw} component and the background model. We use \texttt{cash} statistics \citep{Cash79} and report 90\% confidence intervals from our fits. The results are insensitive to the fit method, fit statistic, and the binning of the data. The resulting light curve of HST-1 is shown in Figure \ref{fig:HST-1_lightcurve}. It is divided into two segments: (i) the period before $\sim$ 2008, where the offset and flux rise and fall, and (ii) the period after $\sim$ 2008, where the offset steadily rises, implying v/c$\sim$6.6$\pm$0.9 (see Section \ref{sec:velocity_calculation}). Below, we discuss each of these periods in turn. 

\subsection{The Flux-Offset Correlation Before 2008}
\label{sec:CTI effect}

One of the most interesting features of HST-1 is the bright X-ray flare detected between 2003 and 2008 \citep{Harris03, Harris06, Harris09}. Before the flare starts, the separation of HST-1 from the core is $\sim$70 pc. During the flare time up to $\sim$2008, the offset is correlated with its flux. Taken at face value, this implies that HST-1 moves away from the core starting in 2002, but then moves back toward the core starting in $\sim$2005. Since this does not seem physically plausible, it raises a question as to whether the apparent motion of the knot can be attributed to an astrophysical process or an instrumental effect. We consider instrumental effects first, focusing on charge transfer inefficiency (CTI) and pileup \citep{Davis01}.

(1) We consider the possibility that CTI could instead be responsible for the flux-offset correlation in HST-1. CTI is the inefficient transfer or loss of charge as it is shifted toward the read-out. The primary effect of CTI is to reduce measured count rates and reduce the ACIS energy resolution, but it can also smear events in the direction of the readout\footnote{\url{https://cxc.cfa.harvard.edu/ciao/why/cti.html\#tech}}. Smearing would tend to shift the centroid of the image and might be more noticeable at higher flux. Since it depends on the relative orientation of the source and the read-out direction, we searched for a roll-angle dependence in our flux-offset correlation. 

\begin{figure}[h!]
	\centering 
	\includegraphics[width=\columnwidth]{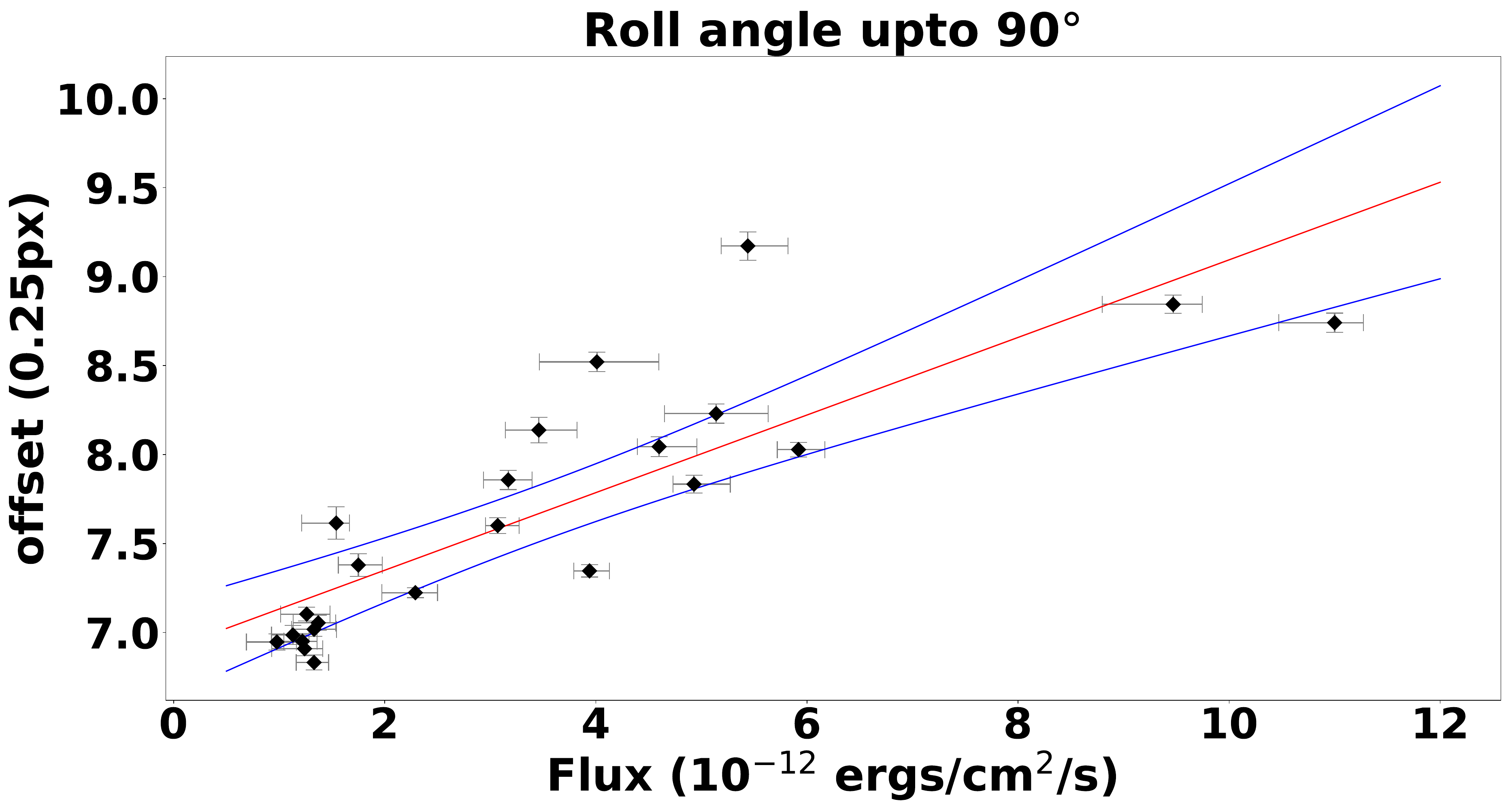}
 	\includegraphics[width=\columnwidth]{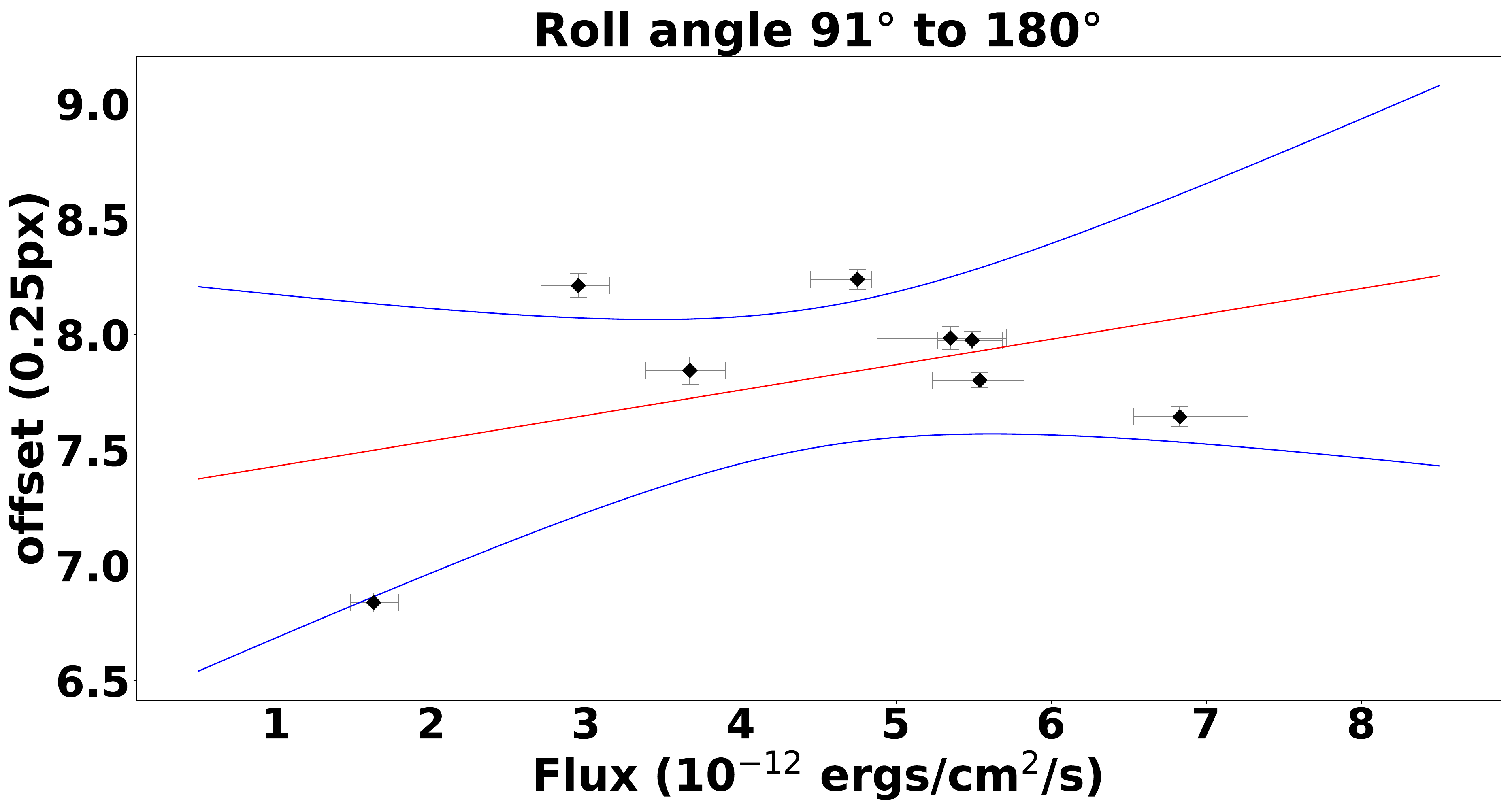}
  	\includegraphics[width=\columnwidth]{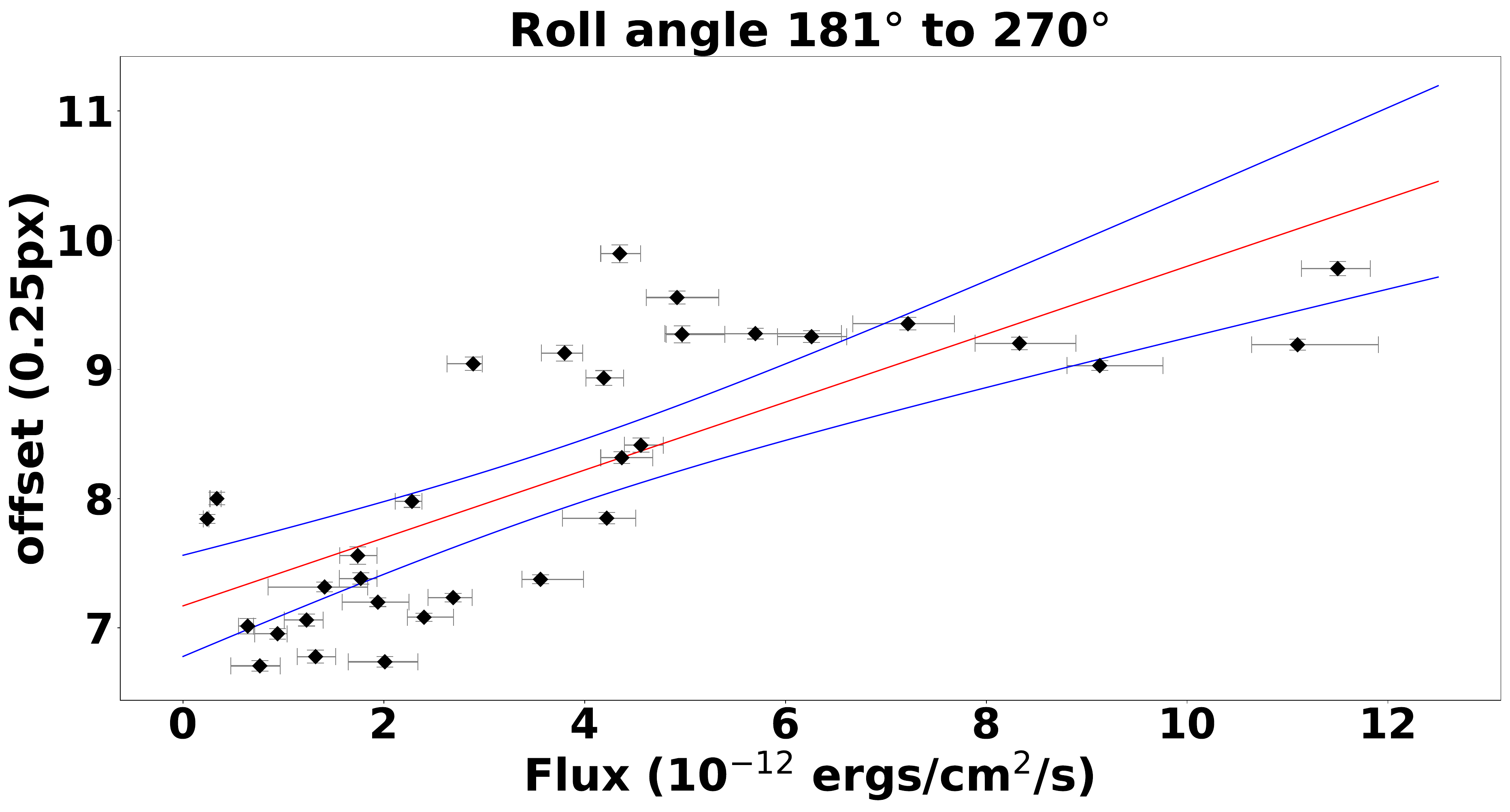}
	\caption{The linear regression model for the flux vs offset of the roll angles (90, 180, 270 degrees) during the period 2000-2008. The model is shown in a red line, while the blue line shows a 95\% confidence region.}
 \label{fig:Roll_angle_90_180_270}
\end{figure}

A roll-angle dependence of the flux-offset correlation could indicate that CTI is driving the pre-2008 behavior of HST-1. To explore this issue, we checked the roll angles (i.e., 0-90$^{\circ}$, 91-180$^{\circ}$, and 181-270$^{\circ}$) of all observations from 2000 to 2008. For each roll angle interval, we performed three linear regression analyses in Python: (1) using {\fontfamily{qcr}\selectfont curve\_fit}, (2) using {\fontfamily{qcr}\selectfont curve\_fit} with errors sampled by Monte Carlo simulation (100000 iterations), and (3) using the {\fontfamily{qcr}\selectfont LINMIX} package, a Bayesian approach to linear regression. Fitted plots are shown in Figure \ref{fig:Roll_angle_90_180_270} for the {\fontfamily{qcr}\selectfont curve\_fit} method. The fitted and simulated values of the slopes are given in Table \ref{tab:fitting_linear}. 

In Table \ref{tab:fitting_linear}, the error bars for the Monte Carlo method are much smaller than the error bars for the other method, which means that the uncertainty in the other fits is primarily due to scatter in the data. If there is no roll angle dependence in the flux-offset correlation, the results in Table 1 should be consistent with a constant slope. Based on the value of the slope for different roll angles, we considered a constant model $M$ in the range $0.01-0.40\times10^{12}$ qp cm$^{2}$s/erg. We then calculated the respective $\chi^2$,
\begin{equation}
\chi^2 = \sum \frac{\left(M - D_i\right)^2}{\sigma_i^2},
\label{eqn:chi2}
\end{equation}
where $D_i$ is the slope of the linear regression model with the {\fontfamily{qcr}\selectfont LINMIX} method for 90$^{\circ}$, 180$^{\circ}$ and 270$^{\circ}$ and $\sigma_i$ is the uncertainty of the slopes. We found that $\chi^2$ is minimized at M=0.23 with a value $\chi^2$=1.13. The corresponding p-value for a $\chi^2$ distribution with two degrees of freedom is 0.56, which indicates that there is no strong evidence of a roll angle dependence in the flux-offset correlation. Therefore we conclude that there is no evidence that CTI plays a significant role in the flux-offset correlation.

\begin{figure}[ht!]
	\centering 
	\includegraphics[width=\columnwidth]{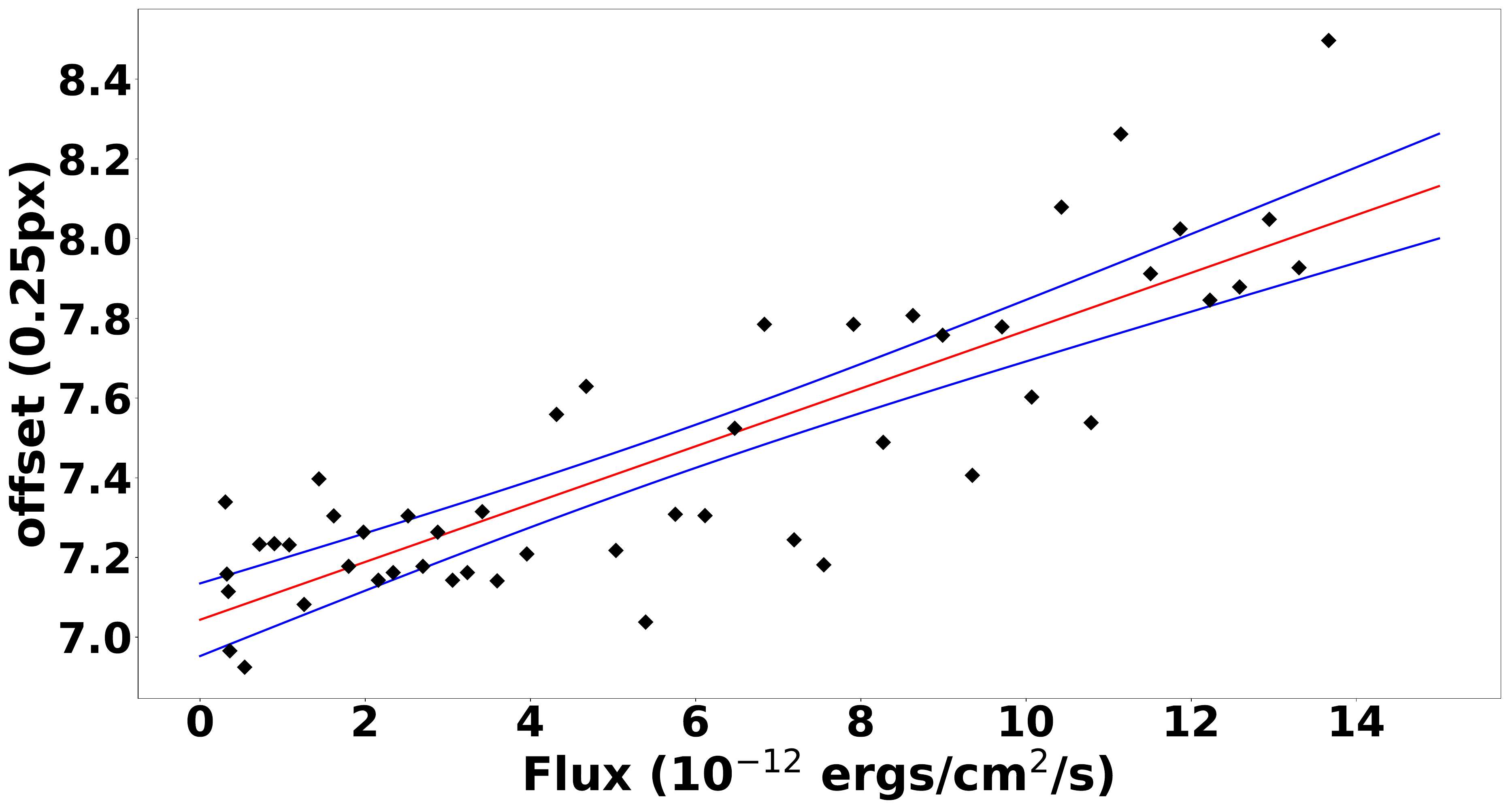}
\caption{Linear regression model for the offset vs simulated flux of HST-1.}
\label{fig:linear}
\end{figure}

(2) When two sources are close together, pileup can occur between them, where their PSFs overlap. The loss of photons between these regions can increase their apparent separation in a way that depends on flux. Therefore, we simulate images of two closely spaced sources (separated by 7\,qp), one at constant flux and one at a wide range of fluxes, to quantify the effect of pileup on the apparent offset between the two sources. We follow the procedure described in Section \ref{sec:analysis} to produce radial profiles of our simulations (accounting for pileup during the {\fontfamily{qcr}\selectfont MARX} simulations), and measured the apparent offsets using 1D Gaussian models. We perform a linear regression analysis for the resulting offset and simulated flux of the HST-1, as shown in Figure \ref{fig:linear}. The slope is 0.073$\pm$0.007 $\times$10$^{12}$qp cm$^{2}$s/erg. We compare this value with our real observational value of HST-1, which at 0.161$\pm$0.015 $\times$10$^{12}$qp cm$^{2}$s/erg is statistically inconsistent with the simulated flux dependence described above. This result shows that the flux-offset correlation in HST-1 is partially but not fully explained by pileup (see Section \ref{sec:C(t)=0+pileup}). Accounting for pileup in our models of the flux-offset correlation is essential for understanding the structure of HST-1. 

\begin{deluxetable}{cccc}[!ht]
\tablecaption{Slope of the flux-offset correlation for the 0-90$^{\circ}$, 91-180$^{\circ}$, and 181-270$^{\circ}$ data sets. Units are quarter pixels per 1$\times10^{-12}$erg/cm$^{2}$s = 1$\times$10$^{12}$qp cm$^{2}$s/erg. Here, qp is a quarter-pixel.} \label{tab:fitting_linear}
\tablehead{\colhead{method} & \colhead{90$^{\circ}$} & \colhead{180$^{\circ}$}  & \colhead{270$^{\circ}$} }
\startdata
Curve fit   & 0.22 $\pm$ 0.03  & 0.11 $\pm$ 0.10 & 0.26 $\pm$ 0.04\\
Monte Carlo & 0.220 $\pm$ 0.006  & 0.109 $\pm$ 0.001 & 0.262 $\pm$ 0.007\\ 
Linmix   & 0.22 $\pm$ 0.03  & 0.11 $\pm$ 0.17 & 0.26 $\pm$ 0.04 
\enddata
\end{deluxetable}

\subsection{The Flux-Offset Model}
\label{sec:Flux-Offset Model}
\begin{figure}[!ht]
	\centering 
	\includegraphics[width=\columnwidth]{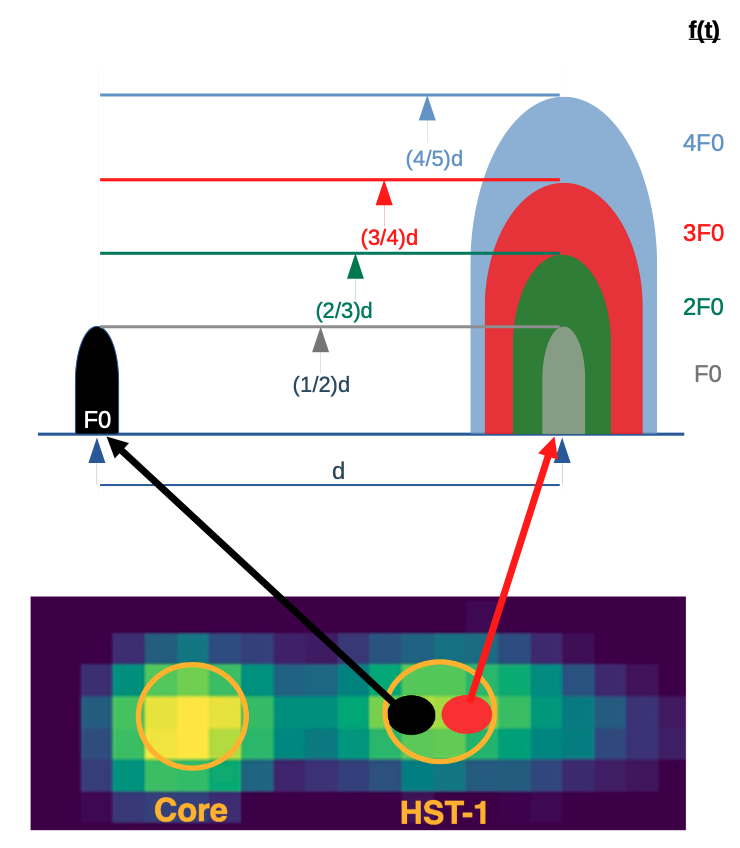}
	\caption{\textit{Bottom panel}: Deconvolved image of core and HST-1 in M87, with two emission regions (black and red circled) shown inside HST-1. \textit{Top panel}: cartoon for Model 1 showing how relative variations in these emitting regions can lead to an apparent shift in the centroid of HST-1 (arrows). This cartoon connects the internal structure of the jet, the separate variation of flux, and the apparent position of HST-1.} 
    \label{fig:HST-1_model}
\end{figure}

Since the apparent motion of HST-1 could not be solely explained by the pileup model (see Section \ref{sec:C(t)=0+pileup}), we consider a toy model to explain why the position of the knot could correlate with flux, appearing to moving away from and then towards the core at highly superluminal speeds. Our model supposes that HST-1 is composed of multiple emitting regions and that the centroid position of the knot reflects their relative fluxes. That is, we assume the position of HST-1 in our data is the flux-weighted centroid of its components. The model is illustrated in Figure \ref{fig:HST-1_model}, containing two unresolved emitting regions: the baseline position of HST-1, with constant flux $F0$ is shown in black, while a variable region with flux $f(t)$ is shown in red a distance $d$ downstream. The arrows show the respective emission regions within HST-1. The top panel is a visualization of the shift of the centroid (whose location is illustrated with a small arrow). As $f(t)$ changes, the apparent position of the knot shifts as well. 

\begin{figure*}[!ht]
	\centering 
	\includegraphics[width=\columnwidth]{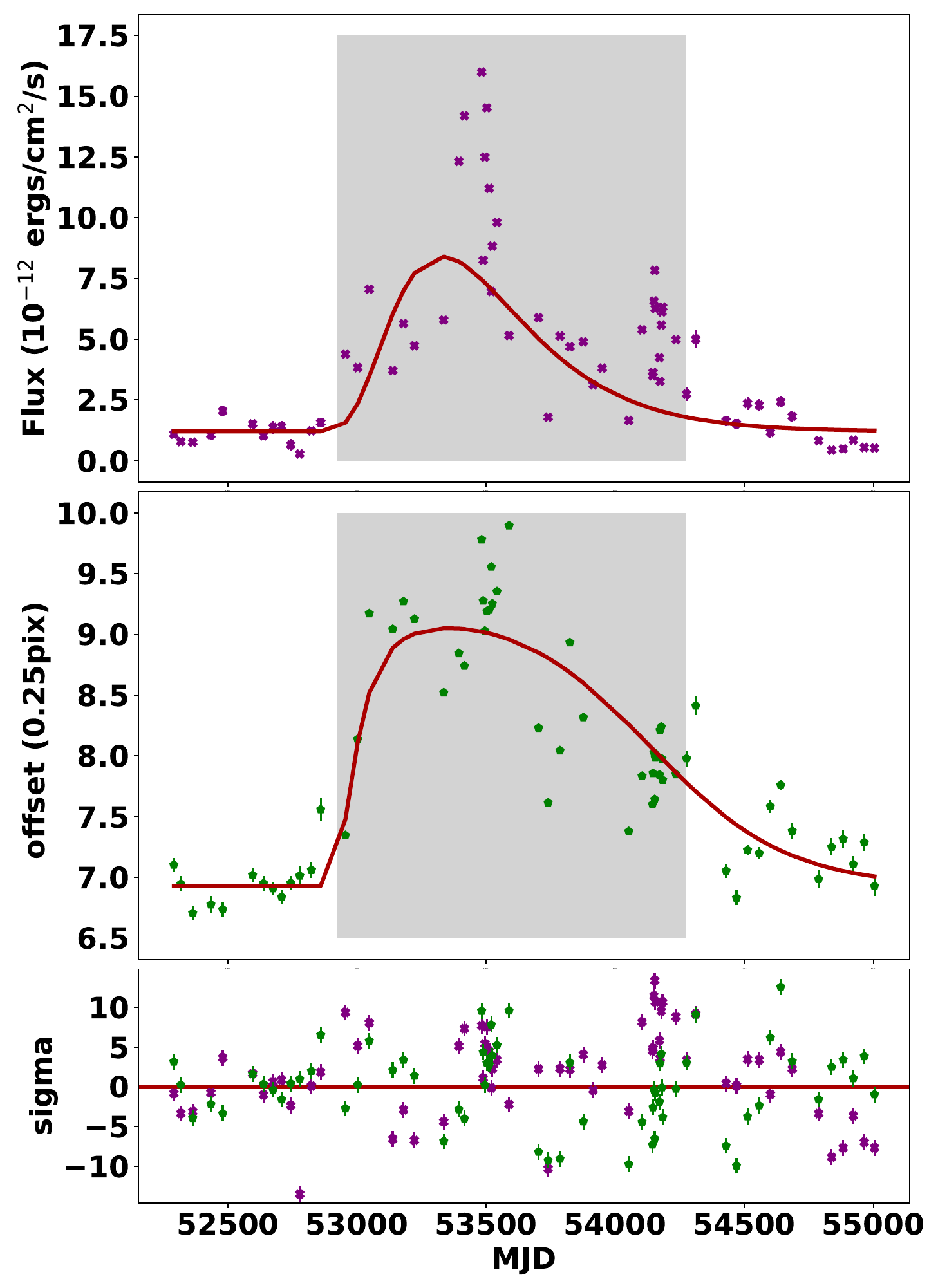}
 	\includegraphics[width=\columnwidth]{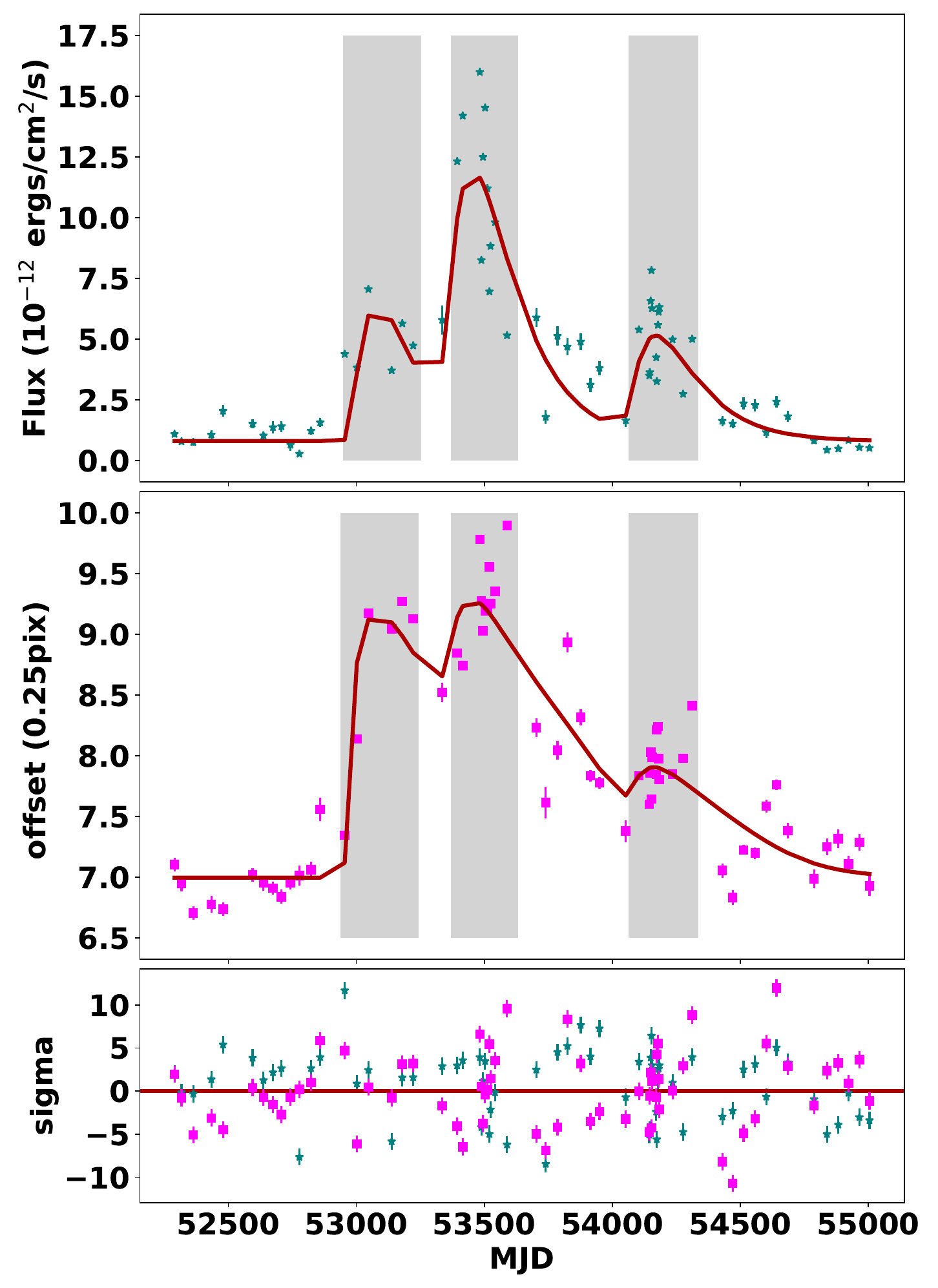}
	\caption{\textit{Left}: Joint fits of Model\,1A (i.e., Eqns. \ref{eqn:model2} and \ref{eqn:model1}) to flux (purple X) and offset (green pentagon) of HST-1. \textit{Right}: Joint fits of Model\,2A to flux (teal star) and offset (magenta square) of HST-1.} 
    \label{fig:model_plot}
\end{figure*}
In this model (i.e., Model 1), the offset $O(t)$ of HST-1 from the core is given by:
\begin{equation}
  O(t) = D + C(t) + a\,f(t), 
   \label{eqn:model1}
\end{equation}
where $D$ is the initial offset of HST-1 before the flare and $C(t)$ is the centroid position, $a$ is the slope value from the flux-offset models to account for pileup, and $f(t)$ is the variable flux. The centroid position is equal to the ratio of the variable flux and total flux times the downstream separation $d$: 
\begin{equation}
 C(t) = \frac{f(t)}{f(t)+F_0} \times [d + V_{ws}(t-t_{o})].
 \label{eqn:model1a}
\end{equation}

\begin{deluxetable*}{ccccccccc}[ht!]
	\tabletypesize{\normalsize}
	\tablecaption{The resulting parameters of the models.} \label{tab:fitting_model}
	
	\tablehead{\colhead{Flares} & \colhead{Parameters} & \colhead{Symbol} 
		& \colhead{Model $1$}  & \colhead{Model $1A$} & \colhead{Model $2$} & \colhead{Model $2A$}  & \colhead{Model $2B$}    & \colhead{Model $2C$} }
	\startdata
	Main flare  &Flux       & F$_{o}$   & 1.08$\pm$0.01 &1.19$\pm$0.02 & 1.00$\pm$0.03 & 0.80$\pm$0.02 & 0.74$\pm$0.03 & 0.85$\pm$0.02 \\
	&Amplitude  & A$_{1}$   & 800$\pm$300   &850$\pm$310   & 660$\pm$700   & 80$\pm$23     & 31$\pm$8      & 28$\pm$7 \\
	&Start Time & t$_{o1}$  & 52800$\pm$20  &52750$\pm$20  & 52800$\pm$40  & 52925$\pm$9  & 52968$\pm$4   & 52975$\pm$3  \\
	&Fall Time  & $\tau_{1}$& 211$\pm$7     &238$\pm$7    & 201 $\pm$20   & 118 $\pm$14   & 130$\pm$16    & 126$\pm$16\\
	&Rise Time  & $\tau_{2}$& 1200$\pm$170  &1300$\pm$160 & 1200$\pm$400  & 208$\pm$30    & 85$\pm$17     & 70$\pm$13 \\
	&Offset     & D  & 7.07$\pm$0.01 &6.84$\pm$0.01 & 6.95$\pm$0.01 & 6.94$\pm$0.01 & 6.93$\pm$0.01 & 6.96$\pm$0.01  \\
	&Distance   & d$_{1}$   & 1.91$_{-0.10}^{+0.02}$ &1.45$\pm$04& 1.83$\pm$ 0.06 & 2.02$\pm$0.06 & 1.75$\pm$0.08 & 2.20$\pm$0.06 \\  
	& Slope      & $a$   & 0 &0.073 & 0 & 0.073 & 0.123$\pm$0.008 & 0.092$\pm$0.008 \\                         
	\hline
	Subflare 1  &Amplitude  & A$_{2}$  &-               &-                  & 30$\pm$5    & 133$\pm$106   & 112$\pm$99          & 106$\pm$59 \\ 
	&Start Time & t$_{o2}$ &-               &-                  & 53400$\pm$4 & 53276$\pm$35  & 53265$\pm$43        & 53288$\pm$25 \\
	&Fall Time  &$\tau_{1}$&-               &-                  & 86$\pm$9    & 145$\pm$13    & 167$\pm$19          & 142$\pm$9 \\
	&Rise Time  &$\tau_{2}$&-               &-                  & 22$\pm$6    & 233$\pm$134   & 251$\pm$160         & 184$\pm$84 \\          
	&Distance   & d$_{2}$  &-               &-                  &2.60$\pm$0.06& 1.56$\pm$0.03 & 1.06$\pm$0.07       & 1.76$\pm$0.00 \\ 
	\hline  
	Subflare 2  &Amplitude  & A$_{3}$  &-               &-                  & 20$\pm$10   & 48$\pm$53     & 40$\pm$42           & 38$\pm$37 \\ 
	&Start Time & t$_{o3}$ &-               &-                  &54000$\pm$50 & 54000$\pm$50  & 54001$\pm$50        & 54006$\pm$44 \\         
	&Fall Time  &$\tau_{1}$&-               &-                  & 150$\pm$20  & 143$\pm$20    & 154$\pm$21          & 144$\pm$18 \\
	&Rise Time  &$\tau_{2}$&-               &-                  & 100$\pm$100 & 214$\pm$180   & 204$\pm$170         & 180$\pm$145 \\
	&Distance   & d$_{3}$  &-               &-                  &0.95$\pm$0.04& 0.66$\pm$0.03 & 0.37$\pm$0.05       & 0.75$\pm$0.02\\ 
	&red.$\chi^2$          & $\chi^2_\nu$              &40.029           & 33.698 &22.940& 20.263 & 20.145     & 20.804
	\enddata
	\tablecomments{Unit of Flux and Amplitude is 10$^{-12}$erg/cm$^{2}$/s; Unit of Start Time, Fall Time and Rise Time is days; Unit of Offset and distance is qp; Slope is 10$^{12}$qp cm$^{2}$s/erg.  
		Model 1 and Model 2 are given before adding $a$ value. Model 1A and Model 2A are the best fit when $a$ is fixed at 0.073. Model 2B is the best fit of Model\,2 when $a$ is set free. Model\,2C is the fit of Model\,2 when $a$ is allowed to vary within its 3$\sigma$ range (see text for details).}
\end{deluxetable*}
The best-fit position of HST-1 is a good approximation to the flux-weighted centroid when $d$ is small (by our estimates $\le$\,1.5 times the FWHM of the deconvolved sources, or $\sim$5 quarter pixels). We consider the upstream region to be stationary, but allow the downstream component to move with velocity $V_{ws}$ after time $t_0$. Since $V_{ws}$ is poorly constrained by our data, we fix it at the previously measured speed of HST-1 (6.4\,c = 5.56$\times$10$^{-4}$qp/day; see section \ref{sec:velocity_calculation} and \cite{Snios19}. This downstream region may be analogous to a working surface in a jet, where ejecta collide and form internal shocks. These have been studied in detail by \cite{Cant00} and \cite{Mendoza09}. These analyses indicate that the lightcurve of a working surface may have a rapid rise and an exponential decay \cite[][Figure 2]{Cant00}. \cite{Coronado16} applied these models in a detailed study of the flaring lightcurve of HST-1. \cite{Nakamura10} developed relativistic magnetohydrodynamic (MHD) models of the shocks in the jet of M87. 

For our fit model for the lightcurve of HST-1, we approximate the exponential rise/decay lightcurve of a working surface as
\begin{equation}
f(t) = A e^{-(\frac{t-t_{o}}{\tau_{1}}+\frac{\tau_{2}}{t-t_{o}})},
\label{eqn:model2}
\end{equation}
where $A$ is the amplitude, $t$ is the time, $\tau_{1}$ and $\tau_{2}$ are fall and rise times. We link the flux and offset models in equations \ref{eqn:model1a} and \ref{eqn:model2} while joint fitting to measure the common parameters using the {\fontfamily{qcr}\selectfont Sherpa} package. Based on these equations we can fit for the separation between two emission regions inside HST-1. The above model is implemented in {\fontfamily{qcr}\selectfont Sherpa} and fitted to the lightcurve and offset of HST-1 as shown in Figure \ref{fig:model_plot} left. The resulting parameters of the models are given in Table \ref{tab:fitting_model}. We used Model 1 to measure the separation between two emission regions inside HST-1 and the initial offset from the core. From Model 1, in the beginning of the flare (MJD 52800$\pm$20) the initial offset from the core is D = 7.07$\pm$0.01 qp $\simeq$ 76.5$\pm$0.1 pc and the separation between the two emitting regions inside HST-1 is d = $1.91_{-0.10}^{+0.02}$ qp $\simeq$ 20.7$_{-1.3}^{+0.2}$ PC. 
For Model 1, we neglect pileup, but we account for it in Model 1A by setting a=0.073 
$\times$10$^{12}$qp cm$^{2}$s/erg in Equation \ref{eqn:model1} (i.e., drawing on our {\fontfamily{qcr}\selectfont MARX} simulations from Section \ref{sec:CTI effect}).
The resulting fit parameters are similar overall, but the initial offset and downstream separation (D and d) are both reduced a little as expected.

In Figure \ref{fig:model_plot} left, the model flux lightcurve increases gradually and then decays exponentially, in broad agreement with the data. The offset model rises suddenly but decays gradually. We attribute the larger scatter in the residuals to scatter in the offset itself. But there are two time periods (MJD 53500 and 54250) where the model does not match the observed flux. This period of time indicates possible additional flares and the discrepancy. This might be solved by adding additional flaring components into the model. 

\subsubsection{Models with pileup and centroid shifts}
\label{sec:C(t)+pileup}
To account for multiple emitting regions (subflares) inside HST-1, we extended our model (i.e., Model 2) of the flux and offset. In Model 2, the flux is the sum of a constant and three flare components while the offset at any time is given by $D$ plus the centroid position of the four emission regions. Model 2 includes additional parameters compared to Model\,1 (7 parameters) as shown in Table \ref{tab:fitting_model}. Similar to Models 1 and 1A, we included the pileup term from Equation \ref{eqn:model1} and set $a$ to 0.073 for Model 2A, to vary freely for Model 2B and to vary within its 3$\sigma$ range for Model 2C. For Model 2A, 2B, and 2C, the parameters are given in Table \ref{tab:fitting_model}. For example in Model 2A: at the beginning of the flare (MJD 52925$\pm$9) the measured initial offset (baseline position) from the core is D = 6.94$\pm$0.01 qp $\simeq$ 75.1$\pm$0.1 pc and the separation between the main flare region and the baseline of HST-1 is d$_{1}$ = 2.02$\pm$0.06 qp $\simeq$ 21.9$\pm$0.6 pc. The first additional flare (subflare 1) occurs $\sim$ 351 days later, at a distance of d$_{2}$\,=\,1.56$\pm$0.03 qp $\simeq$ 16.9$\pm$0.3 pc from the baseline position of HST-1. The separation is smaller than that of the main flare. The second flare (subflare 2) is closer (d$_{3}$ = 0.66$\pm$0.03 qp $\simeq$ 4.0$\pm$0.3 pc), but takes place later at MJD 54000$\pm$50.

Figure \ref{fig:model_plot} (right) shows Model 2A for comparison to Model 1A. The lightcurve of Model 2A rises at MJD 53500 due to the effect of subflare 1. Later, the model flux rises slightly around MJD 54200 due to subflare 2 and then comes back to the baseline of HST-1 by MJD 55000. The offset model curve is similar to the curve in Model 1A, with additional increases in offset due to the effect of subflare\,1 and subflare\,2. Among these three models, we prefer Model 2A, which has the most realistic value of a given our {\fontfamily{qcr}\selectfont MARX}  simulations in Section \ref{sec:CTI effect}.

\subsubsection{Pileup-only models}
\label{sec:C(t)=0+pileup}
Based on the likely contribution of pileup to the offset, we also consider pileup-only models, where the centroid position C(t)=0 appears in Equation \ref{eqn:model1}. In the pileup-only model, the value of $a$ determines how the offset changes with the flux of HST-1. As with Model 2, we set the pileup term $a$ to 0.073 for Model 3A, to vary freely for Model 3B and to vary within its 3$\sigma$ range for Model 3C. Models 3A and 3C do not fit the data nearly as well as models with centroid shifts, but Model 3B provides the best overall fit with reduced $\chi^2$=20.070. The best-fit pileup parameter in this model is $a$=0.292$\pm$0.006$\times$10$^{12}$qp cm$^{2}$s/erg. However, such a strong pileup effect is difficult to reproduce in {\fontfamily{qcr}\selectfont MARX} simulations.
In principle, $a$ might be larger for sources that are closer together and have more PSF overlap between them. Indeed, when we decrease the separation to 5\,qp in our {\fontfamily{qcr}\selectfont MARX} simulations from Section \ref{sec:CTI effect}, we find that $a$ increases to 0.087$\pm$0.009 $\times$10$^{12}$qp cm$^{2}$s/erg. However, this increase appears to be bounded: when we decrease the separation to 3 qp, the value of $a$ drops by a factor of 10 to 0.008$\pm$.011$\times$10$^{12}$qp cm$^{2}$s/erg (likely because the two sources effectively merge into a single source). Therefore, we cannot reasonably attribute such a large value $a$=0.292$\pm$0.006$\times$10$^{12}$qp cm$^{2}$s/erg from Model 3B to pileup between the core and HST-1. Hence, we believe that the pilep-only Model 3B, despite providing the best fit to the data, is unphysical, and we reject it in favor of Model 2A.

\subsection{Superluminal Motion After 2008}
\label{sec:velocity_calculation}

\begin{figure}[!ht]
	\centering 
 	\includegraphics[width=\columnwidth]{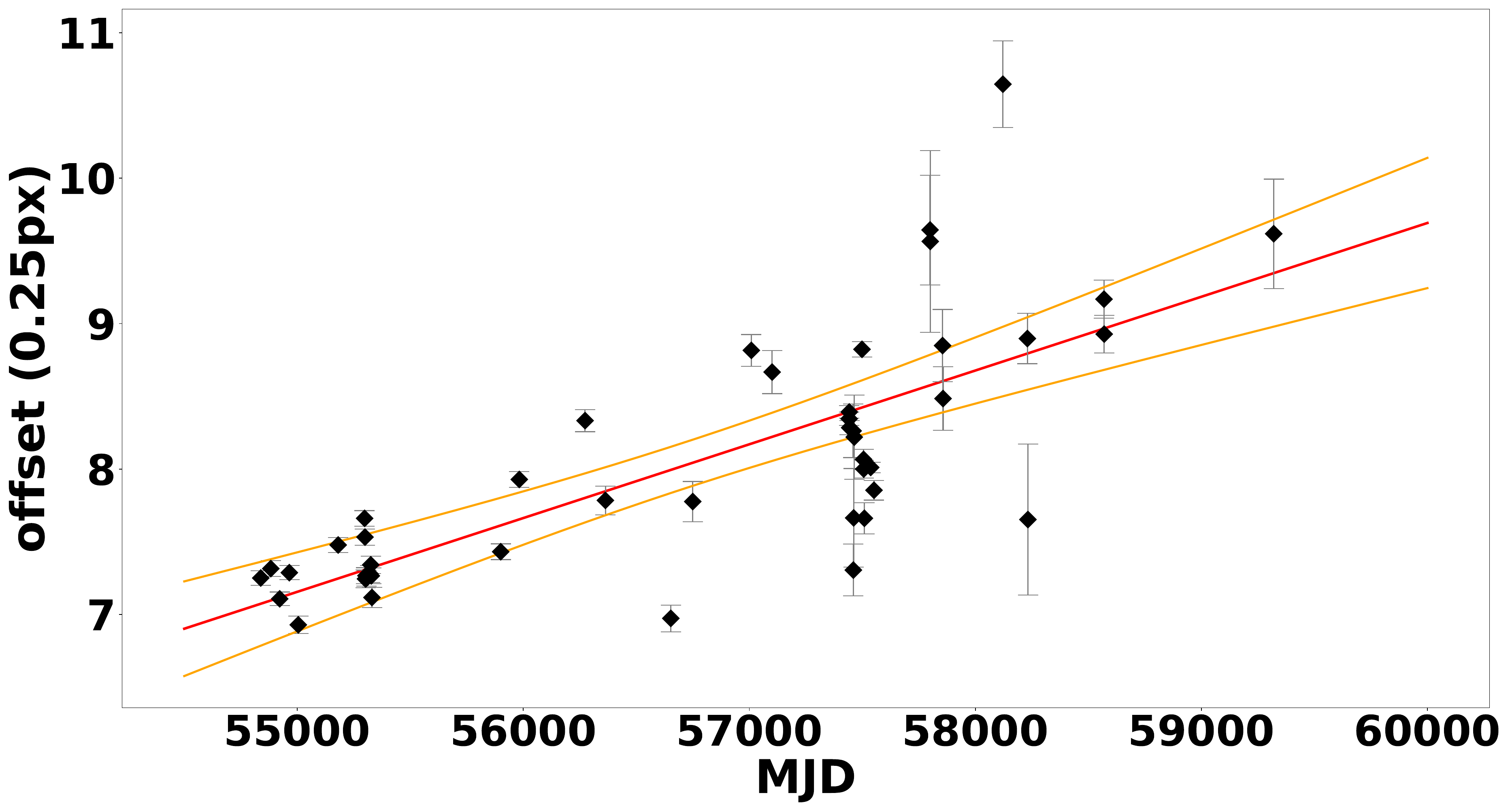}
	\caption{Linear regression model for the offset of HST-1 vs time during the period between 2009-2021 (see Section \ref{sec:velocity_calculation}).}
   \label{fig:speed_v/c}
\end{figure}

After 2008, we found no correlation between the offset and the flux. Instead, there is a steady increase in the offset from 2009 to 2021, see in Figure \ref{fig:speed_v/c}.  We performed a linear regression analysis for the offset vs time in Python using the {\fontfamily{qcr}\selectfont curve\_fit} method. With this model, we measured the bulk speed (v/c) of HST-1. The slope is 0.00051$\pm$0.00007 (qp/day) or 1.3$\pm$0.2$\times$ 10$^{-4}$ pix/day, which corresponds to a measured superluminal speed of 6.6$\pm$0.9 c or 2.0$\pm$0.3 pc yr$^{-1}$. Our measured value is consistent with previously measured values of the velocity from \textit{Hubble} and \textit{Chandra} (\citealp[e.g.,][]{Biretta95, Biretta99, Meyer13, Snios19}). 

\section{Discussion and Conclusion} 
\label{sec: discussion and conclusion} 

We have collected 112-pointings of {\it Chandra} ACIS-S data for the M87 radio galaxy between 2000 and 2021 with a total exposure time of $\sim$ 887 ks and used them to perform image analysis of HST-1. We use astrometric and pile-up corrections and PSF/MARX simulations to produce 0.25-pix deconvolved images. The resulting deconvolved images are used to study the radial profile of the X-ray jet and to measure the relative separation between the core and HST-1. Any proper motion of HST-1 should produce a shift from the core. Our novel relative astrometry makes it possible to measure the proper motion of HST-1 in a large ACIS dataset, even when the absolute astrometry of ACIS is not sufficient for the task \citep{Snios19}. 

Before 2008, we find a correlation between the flux of HST-1 and its offset from the core (see Figure \ref{fig:HST-1_lightcurve} from MJD 53000 to 55000), which coincides with a bright flare from HST-1. As the flare flux rises and falls, HST-1 appears to recede from and then approach the core. We developed a toy model to describe both offset and flux that accounts for physical changes in flux and position of the knot as well as pileup (Section \ref{sec:CTI effect}).

In our model, we attribute changes in flux to additional variable emission regions inside the jet, similar to the working surfaces described in, e.g., \citet{Cant00, Mendoza09, Cabrera13, Coronado16}. As these variable fluxes change, the apparent centroid of HST-1 appears to shift.

Mathematical models were first applied to knots in relativistic jets by \cite{Rees66}, resulting in the measurement of superluminal motion. Since then, analytical and numerical methods have been applied to the internal working surfaces in jets for different sources. \cite{Cant00} derived the injection velocity, density, and momentum in hypersonic jets. \cite{Raga90} measured the variable velocity of an internal working surface to study the proper motion along the stellar jets in variable sources. \cite{Mendoza09} applied a semi-analytical model and measured the luminosity and the working surface velocity from the lightcurve of jets in different GRBs. Later, the same analytical model was applied to the working surface of the blazar PKS 1510-089 to measure physical parameters of the outburst \citep{Cabrera13}. \cite{Coronado16} measured the outburst parameters of HST-1 in the jet of M87 using an internal shock model. In this model, the luminosity in the working surface as a function of time gives an exponential curve. \cite{Nakamura10} developed MHD models of the shocks and applied them to HST-1, consistent with our results.

Based on these studies, we assume an exponential form in Equation \ref{eqn:model2} for the flux lightcurve (see Figure \ref{fig:HST-1_lightcurve}). Our Model\,1A suggests that HST-1 may consist of two emitting regions separated by d = $1.45\pm0.04$ qp $\simeq$ 15.7$\pm0.4$ pc projected distance during the flare. 

Our Model 1A is in broad agreement with the data and covers the overall behavior of the lightcurve and the offset, but there are some limitations to the model. While fitting the model to the lightcurve and offset of HST-1, we observed large residuals which are due to substructure in the flare lightcurve and scatter in the offset. We assumed two emission regions inside HST-1 (i.e., one stable and another moving and variable). But more complex assumptions might fit the data better. For example, we assumed the baseline emission region is a fixed distance $D$ from the core, but its position could change over time. The baseline emission component had constant flux but allowing this component to vary would add additional flexibility to the lightcurve model. One moving, variable emission region but there could be more variable emission regions along the jet. Indeed, \cite{Coronado16} modeled the flares in the lightcurve of HST-1 by considering many working surfaces in the jet of M87.

For Model 2A, we added two flaring emission components to Model 1A to represent two substructures in the lightcurve and the offset. These two regions are assumed to move away from the core at the same speed of V$_{ws}$. Given the success of Model 2A, which fits the data better than Model 1A, it seems that an offset model based on the 100s of internal shocks in \cite{Coronado16}, could fit the data even better. But for our study on the proper motion of HST-1 and to measure the separation between multiple emission regions inside the knot, our simplified models are sufficient.

After 2008, there is a steady increase in the offset of HST-1 and a steady decrease in flux, but the changes in the flux are small compared to the flare. We measured a bulk speed v/c = 6.6$\pm$0.9 or v = 2.0$\pm$0.3 pc yr$^{-1}$ which, is consistent with v = ($6.3\pm0.4$) c as measured by \cite{Snios19} using \textit{Chandra's} HRC-I instrument. The strong consistency validates our relative astrometry. Furthermore, results are in agreement with the previous measurements of \citealp[e.g.,][]{Biretta95, Biretta99, Cheung07, Meyer13}. The superluminal motion is a major study of many other jets in the MOJAVE program and is consistent with our results \citep[i.e., $\sim $5\,c,][]{Cohen07, Homan09}.

Overall, studying the proper motion of jets with high-resolution X-ray observations offers a wealth of knowledge about the dynamics, the process of jet formation, and interaction with the surrounding environment. A more detailed analysis based on the analytical and numerical models above could provide valuable constraints on the particle acceleration mechanisms within the jets, leading to a deeper understanding of the underlying physics of jets from black holes. \\

\hspace{2.5cm}
\textbf{Acknowledgements}

This research has made use of data obtained from the \textit{Chandra} Data Archive. This work was supported by NASA award 80NSSC20K0645 (R.T., J.N). The authors thank the referee for the valuable comments on the manuscript and for pointing out the asymmetric effect of pileup. R.T. and J.N. thank {\L}.\, Stawarz for useful suggestions and discussions on this project.\\


\textit{Software:} CIAO v4.14 \citep{Fruscione06}, Sherpa v4.9.7 \citep{Freeman01}

\bibliography{ms}

\end{document}